\documentclass[doublecol]{epl2} 
\usepackage[pdftex]{}

\title{Pairing Glue Activation in Cuprates within the Quantum Critical Regime}
\shorttitle{Pairing Glue Activation in Cuprates} 

\author{Josef Ashkenazi\inst{1} \and Neil F. Johnson\inst{1}}
\shortauthor{Josef Ashkenazi \etal}

\institute{                    
  \inst{1} Physics Department, University of Miami, P.O.~Box 248046, Coral Gables, FL 33126, U.S.A.
}

\abstract{
A grand challenge in many-body quantum physics is to explain the apparent connection between quantum criticality and high-temperature superconductivity in the cuprates and similar systems, such as the iron pnictides and chalcogenides. Here we argue that the quantum-critical regime plays an essential role in activating a strong-pairing mechanism: although pairing bosons create a symmetry-breaking instability which suppresses pairing, the combination of these broken-symmetry states within the critical regime can {\em restore} this symmetry for the paired quasiparticles. This condition is shown to be met within a large-$U$ ansatz. A hidden quantum phase transition then arises between a Fermi-liquid and a non-Fermi-liquid broken-symmetry striped state, and a critical regime in which the broken-symmetry states are combined. 
}

\begin{document}

\maketitle

\section{Introduction}
Efforts to raise the superconducting (SC) transition temperature $T_c$ in electron--phonon systems, have been hampered by the nature -- and our understanding -- of symmetry-breaking structural instabilities \cite{Testardi} when the electron--phonon coupling constants become too strong. These instabilities open gaps on the Fermi surface (FS), thereby reducing the magnitude of the pairing-strength parameter $\lambda$ \cite{Kresin} and preventing possible increases in $T_c$. Non-phonon mechanisms suffer from a similar problem, i.e., competition between pairing and symmetry-breaking instabilities of spin or charge inhomogeneity. The breakdown of time-reversal symmetry (in the case of a magnetic inhomogeneity) introduces \cite{Kuper1,Kuper2} a further limitation to high-$T_c$ superconductivity (HTSC). 

At the same time, a wide class of unconventional SC systems such as high-$T_c$ cuprates, Fe-based superconductors (FeSCs) and other SC systems, are characterized by anomalous features which are inconsistent with normal Fermi-liquid (FL) behavior. One major anomalous feature is the {\em linear} dependence of the scattering rates on temperature ($T$) or energy ($\omega$), which has been suggested \cite{Varma} as characterizing a non-FL state, known as the ``marginal FL'' (MFL) state. This state has been attributed to quantum criticality \cite{Castro, Sachdev, Zaanen} which occurs in magnetic insulators \cite{Sachdev}, and is characterized by a system-variation parameter $p$. For $p>p_c$ the ground state is of a homogeneous state, and for $p<p_c$ it is of an ordered magnetic state (e.g. a N\'eel antiferromagnetic (AF) state in TlCuCl$_3$ \cite{Sachdev}) which breaks the symmetry between different possible spin arrangements and directions. At the quantum critical point $(p,T) = (p_c,0)$, and a $T>0$ critical regime \cite{Sachdev}, the quantum state is a complex combination of states with macroscopically different symmetries. Adapting such a phase diagram to the cuprates, $p$ becomes the doping level. 

The connection between HTSC and the normal-state properties \cite{Deutscher} remains unclear. Two pictures have been presented: a ``top--down'' approach attributes HTSC to the nature of the normal-state properties, while a ``bottom--up'' approach attributes {\it both} HTSC and the anomalous normal-state properties to quantum criticality. For the second, the existence of a ``critical glue'' has been suggested \cite{Zaanen2} which provides SC pairing. Here we present a {\em hybrid} proposal: while the interactions necessary for strong pairing do exist in the normal state, for low and intermediate values of $p$, the pairing glue is suppressed by competing symmetry-breaking instabilities, {\em unless} there exists a quantum critical regime in which the symmetry-breaking states are combined. In the high-$p$ SC regime our approach could be regarded as a bottom-up approach. 

Our proposed theory, which we refer to as the auxiliary Bose condensates (ABC) theory, focuses on hole-doped cuprates, although its basic conclusions are expected to apply to a wider range of SC systems. Although related to a previous work \cite{AshkHam, Ashkenazi}, there is a crucial difference: In Refs. \cite{AshkHam, Ashkenazi}, the stability of the quantum state of the system, which is a combination of states of different symmetries, was attributed to the energy gain in the anomalous normal and paired states. By stark contrast, the present theory attributes the stability of that state to quantum criticality, as discussed above \cite{Sachdev}. This results in several distinct consequences, despite the similar quantum ansatz.

The minimal reduced Hamiltonian necessary to study the major low-energy features of the cuprates is \cite{Ashkenazi} a one-band 2D Hubbard model on a square lattice (unit vectors $a{\hat x}$ and $a{\hat y}$), with transfer integrals up to the third-nearest neighbor (i.e. including $t$, $t^{\prime}$ and $t^{\prime\prime}$). The on-site Coulomb repulsion parameter $U$ within this band corresponds to the large-to-intermediate-$U$ regime. A dynamical cluster quantum Monte Carlo study of the 2D Hubbard model \cite{Jarrell1, Jarrell3} confirmed the existence of a MFL regime, between pseudogap (PG) and FL regimes, but an attempt \cite{Jarrell2} to prove that the calculated $T_c$ is raised within the MFL regime was less conclusive \cite{Jarrell3}. Since it was based on electron-like quasiparticles (QPs) which are paired due to their coupling to spin fluctuations, and such QPs correspond to a FL state, it is doubtful whether a pairing theory in a non-FL state could be based on them. By contrast, QPs appropriate for large-$U$ systems can be determined using the auxiliary-particle approach \cite{Barnes}, and this should also provide a reasonable description in the intermediate-$U$ regime. Auxiliary particles can be described \cite{AshkHam} as combinations of atomic-like configurations with the same number of electrons per site. Hence beside the cuprates, this approach is also applicable to low-energy excitations in multi-band systems such as the FeSCs.

\section{Results}
Within the auxiliary-particles method, we are free to choose configurations corresponding to the number of electrons per site in the undoped ($p = 0$) case to be bosons, which we call svivons. Those corresponding to one electron more, or less, per site are then fermions, called here quasi-electrons (QEs). 
Due to the presence of dynamical inhomogeneities, the auxiliary-particles constraint \cite{AshkHam, Ashkenazi} is maintained in a site-dependent and time-dependent manner. Hence a dynamical field of Lagrange multipliers is introduced which is equivalent to a field of bosons, which we call lagrons. Within a grand-canonical scheme, the Hamiltonian includes QE--lagron and svivon--lagron coupling terms. Due to the nature of the constraint, they represent the coupling of electrons to spin, orbital and charge fluctuations.
Consequently, each of the three above QPs has a distinctive major role in the physics of the system: the svivons are mainly involved in the establishment of spin/orbital/charge inhomogeneities, due to their Bose condensation; the doping-induced fermion QEs are the major QPs behind charge dynamics; and the lagrons play the role of the scattering/pairing bosons.

Studies \cite{Emery1} using similar Hamiltonians predict that the interplay between electron hopping and AF exchange in the cuprates can drive the formation of the empirically observed striped structures \cite{Tran1, Tran2}, characterized by a spin density wave (SDW) with wave vector:
\begin{equation}
{\bf Q}_m = {\bf Q} + \delta {\bf q}_m,\ {\rm for} \ m = 1\ {\rm or}\  2\ {\rm or} \  3\ {\rm or} \ 4, \ \ \ \label{eq1} 
\end{equation}
where ${\bf Q} = (\pi/a)({\hat x}+{\hat y})$ is the wave vector of the
AF order in the parent (thus $p=0$) compounds, and $\delta {\bf q}_m = \pm \delta q {\hat x} \ {\rm or} \ \pm \delta q {\hat y}$ are modulations around ${\bf Q}$ (typically $\delta q \cong \pi/4a$). 
Similarly to the N\'eel state in TlCuCl$_3$ \cite{Sachdev}, these striped states break the symmetry between different possible spin arrangements and directions. Here they correspond to a Bose-condensed svivon field whose spectrum has a V-shape energy minimum $k_{_{\rm B}} T /\mathcal{O}(N)$ at the points $\pm {\bf Q}_m /2$ \cite{AshkHam, Ashkenazi}. Consequently, gaps open up on a part of the QE and electron FS, similarly to the case of structural instabilities \cite{Testardi}, and also time-reversal symmetry is removed, both countering SC pairing. However \cite{AshkHam, Ashkenazi}, a combination state of different svivon Bose condensates, corresponding to the symmetry-breaking striped states, can yield HTSC. In this state each condensate is not an eigenstate; consequently the energy minima of the svivon spectrum rise above zero and become parabolic. 

Electron states are convoluted QE--svivon states. Hence their normal Green's-function matrix $\underline{\cal G}^d$ (which is diagonal in the ${\bf k}$ representation) is derived \cite{Ashkenazi} from a ``bare'' matrix $\underline{\cal G}^d_0$, due to bubble diagrams of QE and svivon Green's functions, and its ``dressing'' through QE--svivon scattering processes induced by inter-site  transfer which is expressed in terms of a matrix $\underline{\tilde t}$. Multiple-scattering introduces a self-energy correction $\underline{\Sigma}^d \cong \underline{\tilde t} (\underline{1} - \underline{\cal G}^d_0 \underline{\tilde t})^{-1}$, yielding:
\begin{equation} 
\underline{\cal G}^d \cong \big( \underline{1} - \underline{\cal G}^d_0
\underline{\tilde t} \big) \big( \underline{1} - 2\underline{\cal G}^d_0 \underline{\tilde t} \big)^{-1} \underline{\cal G}^d_0. \label{eq6} 
\end{equation}
Consequently, $\underline{\cal G}^d$ has two types of poles: ({\it i}) $\underline{\cal G}^d_0$ introduces a continuity of poles per ${\bf k}$ state, with maximal contribution from svivons close to their energy minima, and thus a non-FL-type feature; ({\it ii}) $(\underline{1} - 2\underline{\cal G}^d_0 \underline{\tilde t})^{-1}$ introduces one pole per ${\bf k}$ state, and thus a FL-type feature. The quantum state of normal-state electrons is therefore a combination of a FL state and non-FL states of different broken-symmetry striped structures, characteristic of a quantum critical regime \cite{Sachdev}. Indeed, the observed critical regime appears projected from a $T=0$ quantum phase transition which is missing in the presence of the SC regime. This hidden transition is between a low-$p$ broken-symmetry non-FL striped state, and a high-$p$ homogeneous FL state. 

Since the broken-symmetry striped states correspond to svivon condensates, their combination restores symmetry {\it vis a vis} the QEs. This restoration is possible here because the degrees of freedom of the electrons are separated between QPs associated with the inhomogeneities (the svivons) and those associated with charge dynamics (the QEs). 
The evaluation of the QE spectrum, and that of the electron, is based on an adiabatic treatment of the combined states. The resulting QE spectrum \cite{Ashkenazi} consists of Brillouin-zone (BZ) areas of flat, polaron like, low-energy QEs, and an abrupt transition between them and BZ areas of high-energy QEs. This feature is reflected in the electron spectrum in the existence of kinks and ``waterfall'' features.

\begin{figure}  
\centering
\includegraphics[width=\columnwidth]{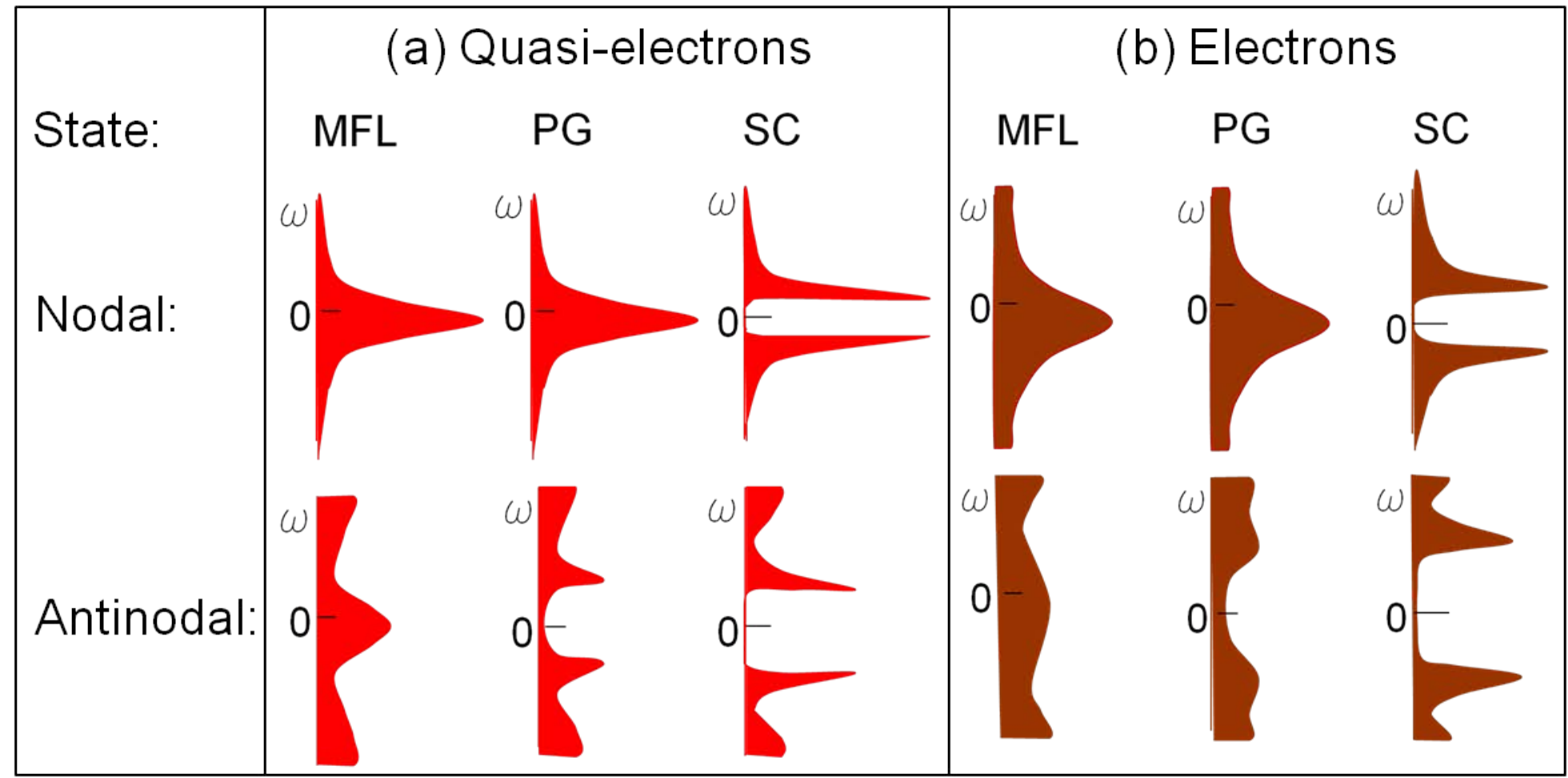}
\caption{Typical low-energy (a) QE and (b) electron spectral functions in the MFL, PG and SC states, at nodal and antinodal points in hole-doped cuprates. The antinodal QEs are paired both in the PG and the SC states, while nodal QEs are paired only in the SC state.}
\label{fig1}
\end{figure}

\begin{figure}  
\centering
\includegraphics[width=\columnwidth]{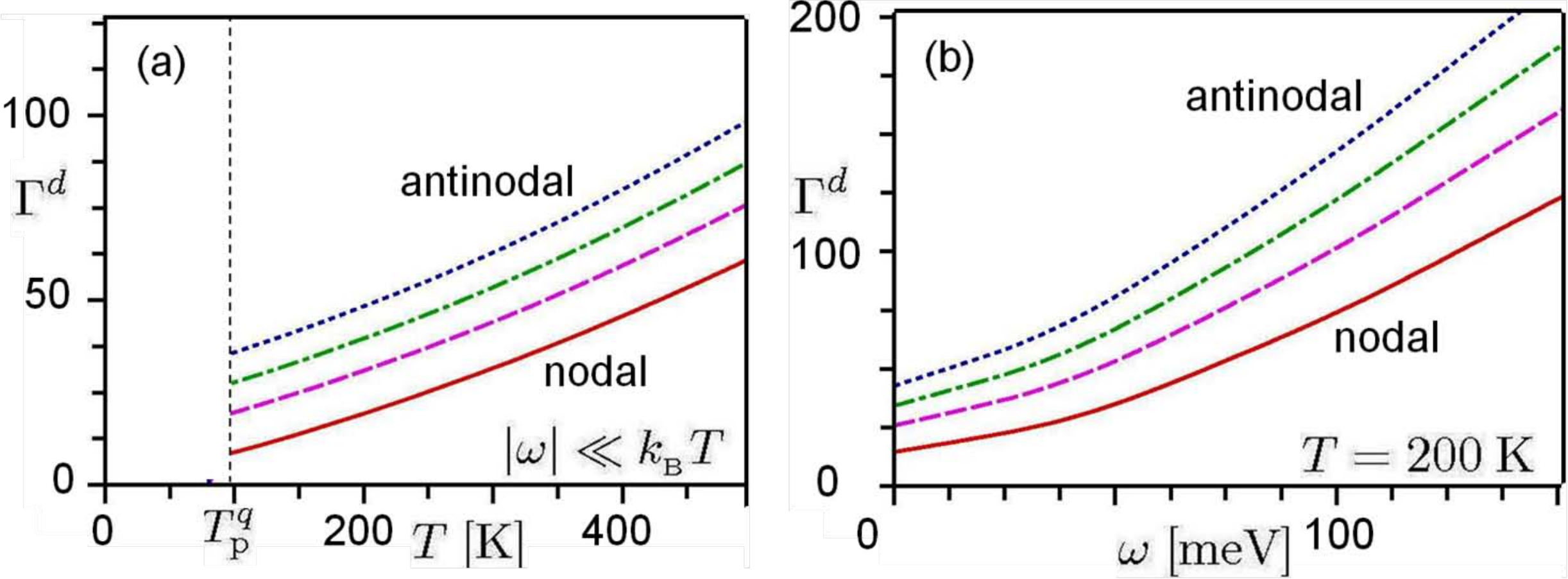}
\caption{Typical trends for (a) the $T$ dependence, and (b) the $\omega$ dependence of the electron scattering rates $\Gamma^d$, within the MFL regime (i.e. $T$ above the QE pairing temperature $T^q_{\rm p}$) at ${\bf k}$ points ranging between the nodal and antinodal BZ areas.}\label{fig2}
\end{figure}

The QE spectral functions within the low-energy BZ areas reflect the effect of the combined states. For a static striped state of one of the ${\bf Q}_m$ wave vectors in Eq.~(\ref{eq1}), a zero-energy gap opens up at low $T$ in ${\bf k}$ points of low-energy QEs, if there are low energy QEs close to ${\bf k} + {\bf Q}_m$ or ${\bf k} - {\bf Q}_m$. This results in a split-peak QE spectral structure near such ${\bf k}$ points, and a one-peak QE spectral structure at other low-energy ${\bf k}$ points. 
Thus, for a combination of striped states of the different ${\bf Q}_m$ wave vectors there are low-QE-energy ${\bf k}$ points close to the antinodal BZ areas, where the split-peak and one-peak scenarios are combined, while at other ${\bf k}$ points close to the nodal BZ areas, the single-peak scenario prevails. 
These spectral features are sketched in Fig.~\ref{fig1}(a). Since each peak in the spectral functions corresponds to a pole in the QE Green's functions, the QE BZ is divided into two analytically distinct $T$-dependent areas where they have one pole per ${\bf k}$ point in one, and three poles in the other. Fig.~\ref{fig1}(b) shows the reflection of these features in the low-energy electron spectral functions.
The results for the electron scattering rates $\Gamma^d({\bf k},\omega,T)$ are presented in Fig.~\ref{fig2}. For low $\omega$ and $T$ these results are approximated by terms of the form: 
\begin{equation}
\Gamma^d({\bf k},\omega,T) \cong \Gamma^d_0({\bf k}) - \Gamma^d_1({\bf k}) |\omega| b_{_T}(-|\omega|),  
\label{eq2} 
\end{equation}
where the Bose distribution function $b_{_T}$ emerges from the QE--svivon convolution expression \cite{Ashkenazi}. Eq.~(\ref{eq2}) yields that $\Gamma^d({\bf k},\omega,T)$ can be approximated as $\Gamma^d_0({\bf k}) + \Gamma^d_1({\bf k}) k_{_{\rm B}}T$, for $|\omega| \ll k_{_{\rm B}}T$, and $\Gamma^d_0({\bf k}) + \Gamma^d_1({\bf k})|\omega|$, for $|\omega| \gg k_{_{\rm B}}T$, in agreement with MFL phenomenology \cite{Varma}.

QE--lagron coupling induces QE and electron pairing. However, the process of coupling of electrons to spin/orbital/charge fluctuations, represented here through svivon--lagron and QE--lagron coupling, results in the striped states, and it causes pairing when these states are combined due to quantum criticality. Hence the lagrons are playing the role of a {\em critical pairing glue} which is activated in the critical regime; it is deactivated in static striped states, and fades away deep within the FL regime.
Since the BZ of the low-energy QEs consists of two analytically distinct parts, pairing can occur below different temperatures, $T^*$ and $T_c$ ($T^* > T_c$) in the antinodal and nodal areas respectively. SC occurs for $T < T_c$, while the PG occurs for $T_c < T < T^*$. Fig.~\ref{fig1} illustrates the effect of pairing on the low-energy QE spectral structure in these two states, and on low-energy electron spectral functions. 

\section{Discussion}
The pairing-strength parameter $\lambda$, due to QE--lagron coupling, within the quantum critical regime, is estimated \cite{Ashkenazi} to be $\lambda \simeq 3$. Such large values explain \cite{Kresin} the enhancement of $T_c$ compared to conventional superconductors, where large values of $\lambda$ are prevented due to lattice instabilities \cite{Testardi}. Partial pairing around the antinodal BZ areas does occur in the PG state, but does not result in Cooper pairs which carry supercurrent (due to scattering between different $({\bf k}, -{\bf k})$ states without breaking). In the PG state, QE pairs are broken due to scattering between the antinodal and the nodal BZ areas. There is mounting evidence that a regime of SC fluctuations generally exists in the cuprates above $T_c$ \cite{Alloul, Johnson}. 

For $T > T^*$, stripe dynamics are fast, and the short lifetime of the striped structures yields features which are too broad to be observed experimentally. This changes when at least a partial gap opens up in the PG and SC states. The existence of fluctuating stripes has been observed for $T < T^*$ using STM \cite{Yazdani}, and the resulting broken rotational symmetry through the anisotropic Nernst effect \cite{Daou}. The existence of fluctuating stripes (Eq.~(\ref{eq1})) implies that experiments measuring timescales shorter than the stripe dynamics will detect violations of time-reversal symmetry, while longer timescale experiments will not. Neutron scattering measurements \cite{Bourges1} indicate such violation for $T < T^*$, while Zeeman-perturbed NQR results \cite{Keller1} do not indicate it. For fluctuating stripes, pairing does not suffer from time-reversal-symmetry violation.
   
\begin{figure}  
\centering
\includegraphics[width=\columnwidth]{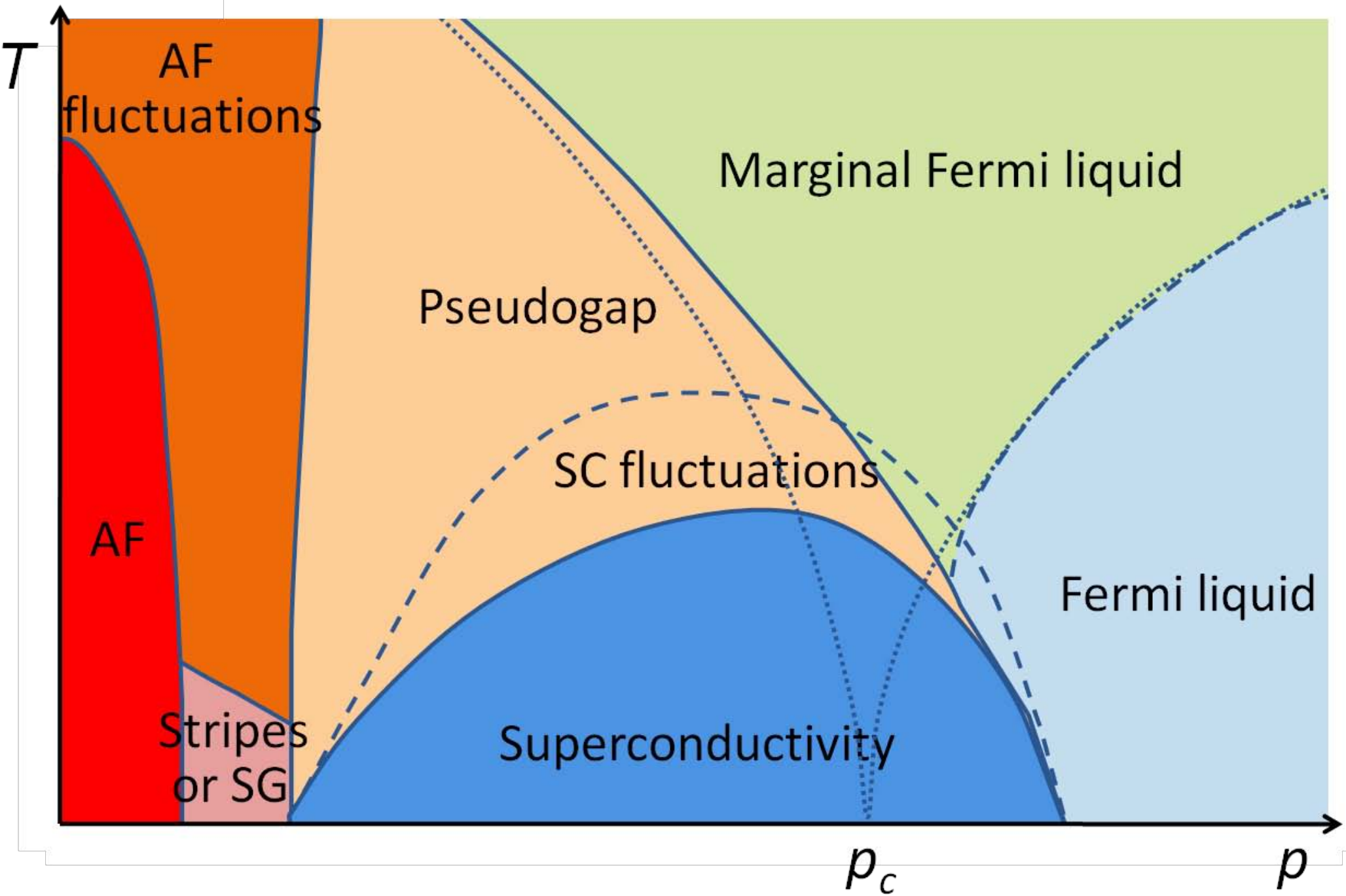}
\caption{Our proposed phase diagram for the hole-doped cuprates, under changing temperature and doping level $p$. Solid and dashed lines represent phase and gradual transitions, respectively. Dotted lines represent gradual transitions in the case that pairing is suppressed down to $T=0$} 
\label{fig3}
\end{figure}

The resulting phase diagram is sketched in Fig.~\ref{fig3}. In the case that pairing is suppressed down to $T=0$, one gets, similarly to magnetic insulators \cite{Sachdev}, a quantum critical point at $(p,T) = (p_c,0)$, and a critical regime between the dotted lines in Fig.~\ref{fig3}. However, the pairing-energy gain due to the establishment of the PG and SC states within the critical regime, {\em extends} the range of existence of this regime (in the sense that the quantum state of the system is a combination including the symmetry-breaking states) {\em into} the PG and SC regimes in Fig.~\ref{fig3}. Consequently, there exists a range of the phase diagram where a direct FL--SC transition occurs through the boundary of the extended critical regime.
The MFL regime is reminiscent of the unextended critical regime, and thus the transition between it and the FL regime remains gradual; it is manifested as a crossover between the linear behavior of the scattering rates (Fig.~\ref{fig2}) and a quadratic behavior characteristic of a FL. Typically for a gradual transition, remnants of MFL behavior persist within the normal state in the overdoped regime \cite{Kokalj}. By contrast, the transition between the MFL and the PG regimes is a phase transition, and our theory predicts that such a transition occurs also between the PG state and the states of lower $p$ (see Fig.~\ref{fig3}). For $p<p_c$, pairing {\it within} the PG and SC regimes can be suppressed under specific doping conditions \cite{Tran1, Tran2}, restoring the broken-symmetry non-FL striped state.

\section{Conclusions}
Similar predictions follow for other SC systems including a quantum-critical regime. The key feature is that the pairing bosons -- which could be of a similar, or a different, nature to the ones in the cuprates -- introduce a symmetry-breaking instability, and these broken-symmetry states are combined in the critical regime, restoring symmetry for the paired QPs. Similar QPs and pairing bosons to those in the cuprates, are expected for the FeSCs, though they will be more evolved due to their multi-band nature. Our theory predicts that the quantum phase transition in the FeSCs is between a non-FL magnetic state and a FL state, and the broken-symmetry states are the perpendicular-direction SDWs \cite{Lynn1} which exist, or almost exist, in the magnetic state.


\begin{thebibliography}{0}
\bibitem{Testardi}L.R.~Testardi, {\it Rev.~Mod.~Phys.} {\bf 47}, 637 (1975).
\bibitem{Kresin}V.Z.~Kresin, and S.A.~Wolf, {\it Rev.~Mod.~Phys.} {\bf 81}, 481 (2009). 
\bibitem{Kuper1}J.~Ashkenazi, C.G.~Kuper, and A.~Ron, {\it Phys.~Rev.~B} {\bf 28}, 418 (1983).
\bibitem{Kuper2}J.~Ashkenazi, C.G.~Kuper, M.~Revzen, A.~Ron, and D.~Schmeltzer, {\it Solid State Commun.} {\bf 51}, 135 (1984).
\bibitem{Varma}C.M.~Varma, P.B.~Littlewood, S.~Schmitt-Rink, E.~Abrahams and A.E.~Ruckenstein, {\it Phys.~Rev.~Lett.} {\bf 63}, 1996 (1989).
\bibitem{Castro}C.~Castellani, C.~Di Castro, and M.~Grilli, {\it Physica C} {\bf 282--287}, 260 (1997).
\bibitem{Sachdev}S. Sachdev and B. Keimer, {\it Physics Today} {\bf 64}, 29 (2011), and references therein.
\bibitem{Zaanen}J. Zaanen, in {\it 100 years of superconductivity}, ed. H.~Rogalla and P.~H.~Kes (Taylor \& Francis, p. 92, 2011).
\bibitem{Deutscher}G.~Deutscher and Y.~Dagan, {\it J.~Supercond.~Nov.~Mag.} {\bf 13}, 699 (2000).
\bibitem{Zaanen2}J.-H.~She, B.J.~Overbosch, Y.-W.~Sun, Y.~Liu, K.~Schalm, J.A.~Mydosh, and J.~Zaanen, {\it Phys.~Rev.~B} {\bf 84}, 144527 (2011).
\bibitem{AshkHam}J.~Ashkenazi, {\it J.~Supercond.~Nov.~Magn.} {\bf 22},
3 (2009).
\bibitem{Ashkenazi}J.~Ashkenazi, {\it J.~Supercond.~Nov.~Mag.} {\bf 24}, 1281 (2011). 
\bibitem{Jarrell1}N.S.~Vidhyadhiraja, A.~Macridin, C.~Sen, M.~Jarrell, and M.~Ma, {\it Phys.~Rev.~Lett.} {\bf 102}, 206407 (2009).
\bibitem{Jarrell3}K.-S.~Chen, S.~Pathak, S.-X.~Yang, S.-Q.~Su, D.~Galanakis, K.~Mikelsons, M.~Jarrell, and J.~Moreno, {\it Phys.~Rev.~B} {\bf 84}, 245107 (2011).
\bibitem{Jarrell2}S.-X.~Yang, H.~Fotso, S.-Q.~Su, D.~Galanakis, E.~Khatami, J.-H.~She, J.~Moreno, J.~Zaanen, and M.~Jarrell1, {\it Phys.~Rev.~Lett.} {\bf 106}, 047004 (2011).
\bibitem{Barnes}S.E.~Barnes, {\it Adv.~Phys.} {\bf 30}, 801 (1981). 
\bibitem{Emery1}V.J.~Emery, and S.A.~Kivelson, {\it Physica C} {\bf
209}, 597 (1993).
\bibitem{Tran1}J.M.~Tranquada, J.D.~Axe, N.~Ichikawa, Y.~Nakamura,
S.~Uchida, and B.~Nachumi, {\it Phys.~Rev.~B} {\bf 54}, 7489 (1996).
\bibitem{Tran2}M.~Fujita, H.~Goka, K.~Yamada, J.M.~Tranquada, and L.P.~Regnault, {\it Phys.~Rev.~B} {\bf 70}, 104517 (2004).
\bibitem{Alloul}H.~Alloul, F.~Rullier-Albenque, B.~Vignolle, D.~Colson, and A.~Forget, {\it Europhys.~Lett.} {\bf 91}, 37005 (2010). 
\bibitem{Johnson}N.F.~Johnson, J.~Ashkenazi, Z.~Zhao, and L.~Quiroga, {\it AIP Advances} {\bf 1}, 012114 (2011), and references therein.
\bibitem{Yazdani}C.V.~Parker, P.~Aynajian, E.H.~da Silva Neto, A.~Pushp, S.~Ono, J.~Wen, Z.~Xu, G.~Gu, and A.~Yazdani, {\it Nature} {\bf 468}, 677 (2010). 
\bibitem{Daou}R.~Daou, J.~Chang, D.~LeBoeuf, O.~Cyr--Choini\`ere, F.~Lalibert\'e, N.~Doiron--Leyraud, B.~J.~Ramshaw, R.~Liang, D.A.~Bonn, W.N.~Hardy, and L.~Taillefer, {\it Nature} {\bf 463}, 519 (2010). 
\bibitem{Bourges1}Y.~Li, V.~Bal\'edent, G.~Yu, N.~Barisi\'c, K.~Hradil, R.A.~Mole, Y.~Sidis, P.~Steffens, X.~Zhao, P.~Bourges, and M.~Greven, {\it Nature} {\bf 468}, 283 (2010).
\bibitem{Keller1}S.~Str\"assle, B.~Graneli, M.~Mali, J.~Roos1, and H.~Keller, {\it Phys.~Rev.~Lett.} {\bf 106}, 097003 (2011). 
\bibitem{Kokalj}J.~Kokalj and R.H.~McKenzie, {\it Phys.~Rev.~Lett.} {\bf 107}, 147001 (2011).
\bibitem{Lynn1} C.~de la Cruz, Q.~Huang, J.W.~Lynn, J.~Li,
W.~Ratcliff II, J.L.~Zarestky, H.A.~Mook, G.F.~Chen, J.L.~Luo,
N.L.~Wang, and P. Dai, {\it Nature} {\bf 453}, 899 (2008). 


\end{thebibliography}
\end{document}